\documentclass[11pt]{article}
\usepackage{rpmacros}
\usepackage[dvipsnames,svgnames,x11names,hyperref]{xcolor}
\usepackage{amsthm}
\usepackage{xfrac}
\usepackage[T1]{fontenc}
\usepackage{enumitem}
\usepackage{mathpazo}
\usepackage{enumitem}
\usepackage{fullpage}
\usepackage{xspace}%
\usepackage[russian,english]{babel}
\allowdisplaybreaks

\usepackage{thmtools, thm-restate}
\usepackage[colorlinks,pagebackref]{hyperref}
\hypersetup{
  linkcolor=[rgb]{0,0,0.4},
  citecolor=[rgb]{0, 0.4, 0},
  urlcolor=[rgb]{0.6, 0, 0}
}
\usepackage{mathrsfs,bbm}
\usepackage{todonotes}
\usepackage{tikz}
\usetikzlibrary{decorations.pathreplacing,backgrounds}
\usepackage{multirow}
\usepackage{setspace}
\usepackage[ruled, linesnumbered, vlined]{algorithm2e}
\SetKwComment{Comment}{/* }{ */}
\SetKwFunction{Prune}{Prune}
\SetKwFunction{FastDESolver}{FastDESolver}
\SetKwFunction{FastFESolver}{FastFESolver}
\SetKwFunction{ShortVector}{ShortVector}
\SetKwFunction{ConstructBasis}{ConstructBasis}
\SetKwFunction{UMultListRecovery}{UMultListRecovery}
\let\oldnl\nl% Store \nl in \oldnl
\newcommand{\nonl}{\renewcommand{\nl}{\let\nl\oldnl}}% Remove line number for one line

\usepackage{amsmath}

\renewcommand{\epsilon}{\varepsilon}
\renewcommand{\tilde}{\widetilde}

\newif\ifdraft
\drafttrue
\ifdraft
\newcommand{\phnote}[1]{\todo[color=red!100!green!33,size=\footnotesize]{ph: #1}}
\newcommand{\sriprahladhuvacha}[1]{\todo[color=red!100!green!33,inline,size=\small]{ph: #1}}
\newcommand{\RGnote}[1]{\textcolor{BrickRed}{\guillemotleft RG: #1 \guillemotright}}
\newcommand{\MKnote}[1]{\textcolor{Orange}{\guillemotleft MK: #1 \guillemotright}}
\newcommand{\ASnote}[1]{\textcolor{Blue}{\guillemotleft AS: #1 \guillemotright}}

\else
\newcommand{\phnote}[1]{}
\newcommand{\sriprahladhuvacha}[1]{}
\newcommand{\RGnote}[1]{}
\newcommand{\MKnote}[1]{}
\newcommand{\ASnote}[1]{}

\fi
\newcommand{\enc}{{\mathsf{Mult}}}

\newcommand{\ehref}[1]{\href{mailto:#1}{#1}}

\newcommand{\vzero}{\mathbf{0}}

\newcommand{\vbeta}{\mathbold{\beta}}
\newcommand{\calB}{\mathcal{B}}
\newcommand{\calU}{\mathcal{U}}
\newcommand{\calV}{\mathcal{V}}
\newcommand{\calW}{\mathcal{W}}

\newcommand{\vv}{\mathbf{v}}

\newcommand{\calM}{\mathcal{M}}
\newcommand{\calL}{\mathcal{L}}
\newcommand{\vB}{\mathbf{B}}
\newcommand{\vR}{\mathbf{R}}
\newcommand{\vU}{\mathbf{U}}
\newcommand{\vV}{\mathbf{V}}
\newcommand{\vW}{\mathbf{W}}

\newcommand{\vY}{\mathbf{Y}}
\newcommand{\vtau}{\mathbold{\tau}}

\newcommand{\ignore}[1]{}

\DeclareMathOperator{\spn}{span}

\usepackage[nameinlink,capitalise]{cleveref}

\usepackage{amsthm}
\usepackage{thmtools,thm-restate}

\numberwithin{equation}{section}
\declaretheoremstyle[bodyfont=\it,qed=\qedsymbol]{noproofstyle}

\declaretheorem[name=Observation,numbered=no]{observation*}

\declaretheorem[numberlike=equation]{theorem}

\declaretheorem[name=Theorem,numbered=no]{theorem*}

\declaretheorem[numberlike=equation]{lemma}
\declaretheorem[name=Lemma,numbered=no]{lemma*}

\declaretheorem[name=Corollary,numbered=no]{corollary*}

\declaretheorem[numberlike=equation]{proposition}
\declaretheorem[name=Proposition,numbered=no]{proposition*}

\declaretheorem[numberlike=equation]{claim}
\declaretheorem[name=Claim,numbered=no]{claim*}

\declaretheorem[name=Conjecture,numbered=no]{conjecture*}

\declaretheorem[name=Question,numbered=no]{question*}

\declaretheoremstyle[bodyfont=\it,qed=$\lrcorner$]{defstyle} 

\declaretheorem[numberlike=equation,style=defstyle]{definition}
\declaretheorem[unnumbered,name=Definition,style=defstyle]{definition*}

\declaretheorem[unnumbered,name=Example,style=defstyle]{example*}

\declaretheorem[unnumbered,name=Notation=defstyle]{notation*}

\declaretheorem[unnumbered,name=Construction,style=defstyle]{construction*}

\declaretheorem[numberlike=equation,style=defstyle]{remark}
\declaretheorem[unnumbered,name=Remark,style=defstyle]{remark*}

\usepackage{nth}
\usepackage{intcalc}
\usepackage{etoolbox}
\usepackage{xstring}

\usepackage{ifpdf}
\ifpdf
\else
\usepackage[quadpoints=false]{hypdvips}
\fi

\newcommand{\ECCCurl}[2]{http://eccc.hpi-web.de/report/\ifnumcomp{#1}{>}{93}{19}{20}#1/#2/}

\newcommand{\shortECCC}[2]{\texttt{\href{http://eccc.hpi-web.de/report/\ifnumcomp{#1}{>}{93}{19}{20}#1/#2/}{eccc:TR#1-#2}}}

\newcommand{\parseECCC}[1]{% Takes a string of the form TRxx/xxx or
\StrSubstitute{#1}{TR}{}[\tmpstring]%
\IfSubStr{\tmpstring}{/}{ %assuming string is of the form TRxx/xxx
\StrBefore{\tmpstring}{/}[\ecccyear]%
\StrBehind{\tmpstring}{/}[\ecccreport]%
}{% assuming string is of the form TRxx-xxx
\StrBefore{\tmpstring}{-}[\ecccyear]%
\StrBehind{\tmpstring}{-}[\ecccreport]%
}%
\shortECCC{\ecccyear}{\ecccreport}}

\title{Fast list recovery of univariate multiplicity and folded Reed-Solomon codes} 
\author{
{Rohan Goyal\thanks{Massachusetts Institute of Technology, Cambridge, MA, United States, Email: \ehref{rohan\_g@mit.edu}. Part of this work was done when the first and second author were visiting the \emph{Simons Institute for the Theory of Computing}, Berkeley. Supported by (Yael Tauman Kalai’s) grant from  Defense Advanced Research Projects Agency
(DARPA) under Contract No. HR0011-25-C-0300. Any opinions, findings and conclusions or recommendations expressed in this material are those of the author(s) and do not necessarily reflect the views of the Defense Advanced Research Projects Agency (DARPA).}}
\and 
{Prahladh Harsha\thanks{Tata Institute of Fundamental Research, Mumbai, India. Email: \href{mailto:prahladh@tifr.res.in}{\{prahladh}, \href{mailto:mrinal@tifr.res.in}{mrinal}, \href{mailto:ashutosh.shankar@tifr.res.in}{ashutosh.shankar\}@tifr.res.in}.  Research supported by the Department of Atomic Energy, Government of India, under project number RTI400112 and partially supported by Google Research Awards, SERB grant and Premji Invest.}}
\and
{Mrinal Kumar\samethanks}
\and
{Ashutosh Shankar\samethanks}}

\date{}

\begin{document}

\maketitle

\abstract{
A recent work of Goyal, Harsha, Kumar and Shankar gave nearly linear time algorithms for the list decoding of Folded Reed-Solomon codes (FRS) and univariate multiplicity codes up to list decoding capacity in their natural setting of parameters. A curious aspect of this work was that unlike most list decoding algorithms for codes that also naturally extend to the problem of list recovery, the algorithm in the work of Goyal et al. seemed to be crucially tied to the problem of list decoding. In particular, it wasn't clear if their algorithm could be generalized to solve the problem of list recovery FRS and univariate multiplicity codes in near linear time. 

In this work, we address this question and design $\tilde{O}(n)$-time algorithms for list recovery of Folded Reed-Solomon codes and univariate Multiplicity codes up to capacity, where $n$ is the blocklength of the code. For our proof, we build upon the lattice based ideas crucially used by Goyal et al. with one additional technical ingredient - we show the construction of appropriately structured lattices over the univariate polynomial ring that \emph{capture} the list recovery problem for these codes.  
}

\newpage

\section{Introduction}

An error-correcting code (or simply a code) $\mathcal{C}$ of block length $n$ over an alphabet $\Sigma$ is said to be $(\rho, L)$ list decodable 
if a Hamming ball of radius $\rho n$ around any vector in $\Sigma^n$ contains at most $L$ codewords from $\mathcal{C}$. When $\rho$ is less than half the minimum distance of the code (denoted by $\delta$), the number of such codewords is at most one, and we are in the so-called unique decoding regime. The notion of list decoding is of great interest both from a combinatorial perspective, where the goal is to understand the tradeoffs between various parameters of the code like $\rho, L, \delta$ and rate as well as from an algorithmic perspective, where the goal is to design efficient algorithms for list decoding the code $\mathcal{C}$ for larger and larger $\rho$. In addition to its considerable inherent interest,  list decodability of codes is also of interest due to deep and extremely fruitful connections to other notions in complexity theory, especially in pseudorandomness \cite{Vadhan2012}. While list decoding was defined in the 1950s by Elias, it is fair to say that the research in this area really took off in the mid-90s following the work of Sudan \cite{Sudan1997} who showed that Reed-Solomon codes can be efficiently list decoded far beyond the unique decoding regime. The subsequent two decades have witnessed significant progress in our understanding of list decoding (cf., \cite{GuruswamiS1999,GuruswamiR2008, GuruswamiW2013,Kopparty2014,KoppartyRSW2023,Srivastava2025,ChenZ2025,JeronimoMST2025,LeviMS2025}) and we now know explicit codes that achieve list decoding capacity with constant list size, and modulo some finer details, this is essentially the best that one can hope for! In fact, we also know that for some of these codes, the list decoding algorithms can be assumed to run in nearly linear time! \cite{GoyalHKS2024, AshvinkumarHS2025, SrivastavaT2025}. 

Our focus in this paper is on the question of list recovery, a notion closely related to that of list decoding that we now define. A code ${\mathcal C} \subseteq \Sigma^n$ is said to be $(\rho, \ell, L)$ list recoverable if for all subsets $E_1, E_2, \ldots, E_n \subseteq \Sigma$ with $|E_i| \leq \ell$, there are at most $L$ codewords $c \in \mathcal{C}$ such that $c_i \notin E_i$ for at most $\rho n$ coordinates $i \in [n]$. Clearly, for $\ell = 1$, this is precisely the notion of list decoding. For a detailed exposition on recent works on list-recovery, we refer to the survey by Resch and Venkitesh \cite{ReschV2025}. In addition to coding theoretic motivations due to its close connection to list decoding, list recovery has found numerous other applications, for instance to the construction of pseudorandom objects like extractors, expanders and condensers \cite{GuruswamiUV2009,KalevT2022}, to cryptography \cite{GoldreichL1989} and to design of algorithms for problems like group testing, compressed sensing \& heavy hitters. We refer the interested reader to detailed expositions of some of these applications in the excellent lecture notes by Mary Wootters \cite{Wootters2019-coding}. 

A curious feature of most of the list decoding algorithms in literature is that they also extend to the problem of list recovery. This includes the algorithms of Sudan \cite{Sudan1997} and Guruswami-Sudan \cite{GuruswamiS1999} for list decoding of Reed-Solomon codes up to the Johnson radius, the algorithms for list decoding Folded Reed-Solomon (FRS) codes, multiplicity codes (both univariate and multivariate versions) \cite{GuruswamiR2008,GuruswamiW2013,Kopparty2014, KoppartyRSW2023,BhandariHKS2024-mgrid,BrakensiekCDZ2025} up to capacity and the recent results on expander based codes \cite{JeronimoMST2025,SrivastavaT2025,JeronimoS2025} that approach list-decoding capacity.  

Yet another family of problems of great interest in the list decoding/recovery setting is to construct codes that can be list decoded/recovered in nearly linear time. This line of work includes results that rely on careful construction of codes that have such super-fast decoding algorithms as well as the design of such algorithms for many known and natural families of codes. A notable milestone in this line of work is the result of Alekhnovich \cite{Alekhnovich2005} that gave near linear time algorithm for list decoding (and recovery) of Reed-Solomon codes up to the Johnson radius. Alekhnovich's algorithm is  essentially a near-linear time implementation of the Guruswami-Sudan algorithm, and introduces and studies some deep connections between the polynomial method-based list decoding algorithms for Reed-Solomon codes in \cite{Sudan1997,GuruswamiS1999} and appropriate lattices over the univariate polynomial ring. In a recent work, Goyal, Harsha, Kumar and Shankar \cite{GoyalHKS2024} built upon the ideas of Alekhnovich to show that Folded Reed-Solomon codes and univariate multiplicity codes codes (which were known to achieve list decoding capacity with constant list size \cite{GuruswamiW2013,Kopparty2014,KoppartyRSW2023,Tamo2024}) can be list decoded up to capacity in nearly linear time. In addition to the lattice-based ideas, this work also gave near-linear time algorithms for solving families of differential equations and functional equations that naturally appear in the context of list decoding and list recovery of FRS and univariate multiplicity codes. 

Given the fact that for most codes, list decoding algorithms also immediately extend to list recovery, it came as a bit of a surprise that the results in \cite{GoyalHKS2024} did not appear to extend to the list recovery problem immediately. Even more surprising was the fact that this already seemed to be the case when the the list size at each coordinate (the parameter $\ell$ in the definition of list recovery) was equal to $2$. In this work, we build upon the techniques of \cite{GoyalHKS2024} to show that the list recovery problem for FRS codes and univariate multiplicity codes can indeed be solved in nearly linear time. More formally, we have the following theorem. 

\begin{restatable}[Main result]{theorem}{mainthm}\label{intro-main} 
For every $\epsilon > 0, \ell \in \N$, there is an $s_0 \in \N$ such that for all $s > s_0$, degree parameter $k$, block length $n$ and field $\F$ of characteristic zero or greater than $k$, the following is true. 

There is a randomized algorithm that when given as input sets $S \subseteq \F$ and $E_{\alpha} \subseteq \F^{s}$ with $|E_{\alpha}| \leq \ell$ for every $\alpha \in S$ and $|S| = n$, runs in time $O\left(n \poly(s, \ell, \log n, \log |\F|, (\ell/\epsilon)^{\frac{1+\log \ell}{\epsilon}}) \right)$ and with high probability outputs the set of all univariate polynomials $f(X) \in \F[X]$ of degree less than $k$ such that for at least $(1 - k/sn - \epsilon)$ fraction of $\alpha \in S$, 
\[
(f(\alpha), f^{(1)}(\alpha), \dots, f^{(s-1)}(\alpha))  \in E_{\alpha} \, .
\]
\end{restatable}

\begin{remark}
  Choosing $s_0$ as $\theta(\ell/\epsilon^2)$ suffices in the above statement.
\end{remark}

\begin{remark}
Even though we state and prove the main theorem for univariate multiplicity codes here, our algorithm extends to Folded Reed-Solomon codes as it is with some cosmetic changes.  
\end{remark}

\subsection{An overview of the proof}
We now give an overview of our proof. We start by setting up some notation. For this overview, we confine ourselves to the list recovery problem for univariate multiplicity codes. 

We have a set $S \subseteq \F$ of size $n$ and the input to the list recovery algorithm is data of the form $(\alpha, \vbeta(\alpha)_{j})_{j \in [\ell], \alpha \in S}$. Here, $\vbeta(\alpha)_j = (\beta(\alpha)_j^{(0)}, \dots, \beta(\alpha)_j^{(s-1)})$ for field elements $\beta(\alpha)_j^{(i)}$. For $\ell = 1$, we are in the list decoding regime. We continue to refer the input as a \emph{received word} even for the list recovery setting for this overview. Before proceeding further, we recall the main steps of the near-linear time list decoding  algorithm in \cite{GoyalHKS2024}, which, in turn is a fast implementation of the algorithm of Guruswami \& Wang \cite{GuruswamiW2013}. The algorithm has two main steps. 
\begin{itemize}
\item \textbf{A differential form that explains the received word: } In the first step the goal is to construct a non-zero polynomial $Q(X, Y_0, \dots, Y_m)$ of the form $Q = \tilde{Q}(X) + \sum_{i} Q_i(X)Y_i$ for an appropriately chosen parameter $m = \Theta(\sqrt{s})$, with the property that the degree of $Q$ is not too large and for every polynomial $f$ whose encoding is close enough to the received word, we have that 
\[
Q(X, f(X), f^{(1)}(X), \dots, f^{(m)}(X)) \equiv 0.
\]
In other words, $Q$ gives us a linear differential equation of high order such that its solution space contains all the close enough codewords. 
\item \textbf{Solving the differential equation: } The second part of the algorithm involves solving the differential equation given by the first part and recovering all close enough codewords. This, in turn, is done in two steps. In the first step, a near-linear time algorithm is designed to obtain a basis of this solution space (which has dimension at most $m$) and then a pruning algorithm of Kopparty, Ron Zewi, Saraf \& Wootters \cite{KoppartyRSW2023,Tamo2024} is invoked to obtain a constant-sized list of close enough codewords. 
\end{itemize}
The high-level structure of the fast list recovery algorithm in this paper also follows the outline above. This should not be surprising since in the work of Guruswami \& Wang \cite{GuruswamiW2013}, an algorithm with the above structure is shown to solve the list recovery problem for univariate multiplicity codes in polynomial time. In fact, beyond the setting of parameters, the list recovery and list decoding algorithms in \cite{GuruswamiW2013} are the same. 

To extend the results in \cite{GoyalHKS2024} to the list recovery setting, we use the fast differential equation solver from \cite{GoyalHKS2024} as it is. However, to extend the first step of constructing the differential equations requires some new technical observations. To elaborate a bit more on these differences, we start by recalling some more of the technical details of the construction of $Q$ in \cite{GoyalHKS2024}. Recall that for the list decoding setting, each $\alpha$ comes with a unique $\vbeta(\alpha)$ in the received word. To construct $Q$, the algorithm in \cite{GoyalHKS2024} starts by constructing univariate polynomials $A_0(X), A_1(X), \ldots, A_{m}(X)$  that satisfy $(A_i)^{(j)}(\alpha) = \beta(\alpha)^{(i + j)}$. These polynomials can be constructed in nearly linear time using standard Fast Fourier Transform-based techniques, quite well known in computational algebra literature. Given these $A_i's$, the authors then look at the set of polynomials in $X, Y_0, Y_1, \dots, Y_m$ generated by taking $\F[X]$-weighted linear combinations of $\{Y_i - A_i(X) : i \in [m]\} \cup \{\prod_{\alpha \in S}(X - \alpha)^{s-m}\}$. This set of polynomials is a lattice over the ring $\F[X]$ (or equivalently, an $\F[X]$-module) and has some nice properties - any non-zero  polynomial of sufficiently low degree in this lattice satisfies the properties desired from $Q$. This follows from the simple observation that if $f(X)$ agrees with the received word at $\alpha$, then for every $i$, $f^{(i)}(X) - A_i(X)$ is zero modulo $(X - \alpha)^{s-i}$. Thus, if we could show that this lattice has a non-zero polynomial of sufficiently low degree and  can also construct one such polynomial in nearly linear time, we would be done. This is shown in \cite{GoyalHKS2024} using an analogue of Minkowski's theorem for this lattice and an algorithm of Alekhnovich for computing \emph{shortest vectors} (under the degree norm) for such lattices. 

The primary difficulty in generalizing this specific approach to the setting of list recovery stems from the fact that in the list recovery setting, we now have many $\vbeta(\alpha)$ for every $\alpha$. Thus, the definition of the the polynomials $A_i$ immediately breaks down since the received word can no longer be seen as a function from the evaluation points $\alpha$ to vectors $\vbeta$. One potential  fix for this situation is to consider the given received word as a union of $\ell$ different received words (with an arbitrary split) in the list decoding setup, and try to work with them together. So, we can try to view the received word $(\alpha, \vbeta(\alpha)_{j})_{j \in [\ell], \alpha \in S}$ as $\ell$ different received words $r_0, \ldots, r_{\ell -1}$ defined as $r_i = (\alpha, \vbeta(\alpha)_{i})_{\alpha \in S}$. Now, for each of these $r_i$, we follow the strategy in the list decoding algorithm and construct the polynomials $A_{i,0}, \dots, A_{i,m}$. To consolidate this information, we could now choose to work over the lattice generated by the polynomials $\prod_{i = 0}^{\ell-1} (Y_j - A_{i, j})$ over the ring $\F[X]$. Even if we  succeeded in finding a low degree non-zero polynomial $Q$ in this new lattice (in near linear time) and showing that it provides us with a differential equation satisfied by all close enough codewords, we have a technical issue to resolve in the second step: the differential equation at hand is no longer linear and it is unclear how to go about solving it and recovering close enough codewords. The solution space of such a differential equation is not necessarily an affine space, and the techniques in \cite{GoyalHKS2024} don't seem to be immediately applicable. We note that in certain coding-theoretic contexts (e.g. \cite{Sudan1997, GuruswamiS1999}) we do indeed work with polynomials $Q$ with high $Y$-degree, although for the particular situation at hand, we do not know how to proceed with this approach. 

Given the technical difficulties outlined above, it seems natural to try and find a $Q$ that is linear in the $Y$-variables for the interpolation step even in the list recovery setting. Guruswami \& Wang show that this can indeed be done in polynomial time. For our proof, we show that this can in fact be done in nearly linear time. To this end, we give a careful construction of a lattice whose generators are linear in the $Y$-variables, such that the shortest non-zero vectors in this lattice are precisely differential equations that \emph{capture} all close enough codewords. See \autoref{sec:technical} for details. 

\subsection*{Organization}
The paper is organized as follows. \Cref{sec:prelims} contains preliminaries related to the univariate multiplicity code and the subroutines invoked during the lattice construction. \Cref{sec:technical} contains the technical details for construction of the differential equation, via fast construction of the previously mentioned lattice. In \cref{sec:proof of main thm}, we mention the subroutines used to extract the list of codewords from the differential equation and give the final proof of \cref{intro-main}. 

\section{Preliminaries}\label{sec:prelims}

\subsection{Notation}

\begin{itemize}
  \item We work in some fixed finite field $\F$. 
  \item For a natural number $N$, we will use $[N]$ to refer to the set $\{0, 1, 2, \dotsc, N-1\}$.
  \item We use $\vzero_k$ to denote the $k$-length vector of all-zeroes. We may drop the subscript if the length is clear from context.
\end{itemize}

\subsection{Univariate multiplicity codes}

The following is the formal definition of the univariate multiplicity code.

\begin{definition}[Multiplicity codes]\label{def:multiplicity codes}
Let $n, k, s$ be positive integers satisfying $ks < n$, $\F$ be a field of characteristic zero or at least $k$ and size at least $n$, and let $S=\{\alpha_1, \alpha_2, \ldots, \alpha_n\}$ be a set of $n$ distinct elements of $\F$. Then, univariate multiplicity codes with multiplicity parameter $s$ for degree-$k$ polynomials is defined as follows. 

The alphabet of the code is $E = \F^{s}$ and the block length is $n=|S|$, where the coordinates of a codeword are indexed by $\alpha \in S$. The message space is the set of all univariate polynomials of degree at most $k$ in $\F[x]$. And, the encoding of such a message $f \in \F[x]$, denoted by $\enc_{s}(f) \in E^S$ is defined as 
\[
\enc_{s}(f)|_{\alpha} := \left(f(\alpha), f^{(1)}(\alpha), \ldots, f^{(s-1)}(\alpha)\right) \, ,
\]
where $f^{(j)}(\alpha)$ is the evaluation of the $j^{th}$-order derivative\footnote{\label{fn:mult-ideal-defn}This definition in terms of (standard) derivatives requires the characteristic of the field to be 0 or at least $d$. All known capacity-achieving list-decoding results for univariate multiplicity codes require large characteristic (previous works \cite{Kopparty2014,GuruswamiW2013,KoppartyRSW2023,GoyalHKS2024} 
as well as our work). For this reason as well as ease of presentation, we work with standard derivatives.} of $f$ on the input $\alpha$. 
\end{definition}

In the list recovery setting, the received word $R$ consists of: for every evaluation point $\alpha \in S$, a collection of tuples $\vbeta(\alpha)_j = (\beta(\alpha)_j^{(0)}, \beta(\alpha)_j^{(1)}, \dotsc, \beta(\alpha)_j^{(s-1)})$ for every $j \in [\ell]$. In words $\beta(\alpha)_j^{(i)}$ is the $j^{th}$ option for the $i^{th}$ derivative of the message polynomial at $\alpha$. Agreement is with respect to inclusion in this list; that is, $\enc(f)$ agrees with $R$ on point $\alpha$ if there exists $j \in [\ell]$ such that $(f(\alpha), f^{(1)}(\alpha), \ldots, f^{(s-1)}(\alpha)) = \vbeta(\alpha)_j=(\beta(\alpha)_j^{(0)}, \beta(\alpha)_j^{(1)}, \dotsc, \beta(\alpha)_j^{(s-1)})$.

Our goal is to carry out list recovery from an agreement fraction of $k/ns + \epsilon$, for a given parameter $\epsilon$.

\subsection{Polynomial operators and lattices}

We recall below the $\tau$ operator, used in \cite{GuruswamiW2013} and \cite{GoyalHKS2024}.

\begin{definition}[Tau operator]\label{def: tau operator}
Let $s \in \N$ be a parameter. Then, for every $m$ such that $0 \leq m < s$, the function $\tau$ is an $\F$-linear map from the set of polynomials in $\F[X, Y_0, Y_1, \ldots, Y_m]$ with $\vY$-degree $1$ to the set of polynomials in $\F[X, Y_0, Y_1, \ldots, Y_{m+1}]$ with $\vY$-degree $1$ and is defined as follows. 
\[
\tau\left(\tilde{Q}(X) + \sum_{i = 0}^m Q_i(X)\cdot Y_i\right) := \tilde{Q}^{(1)}(X) + \sum_{i = 0}^m \left( Q_i^{(1)}(X) \cdot Y_i+ Q_i(X)\cdot Y_{i+1}\right) \, .
\]

For an integer $i \leq (s-m) $, we use $\tau^{(i)}$ to denote the linear map from $\F[X, Y_0, Y_1, \ldots, Y_m]$ to $\F[X, Y_0, Y_1, \ldots, Y_{m+1}]$ obtained by applying the operator $\tau$ iteratively $i$ times.
\end{definition}

The operator is defined this way so that 
\[ \frac{d}{dX} Q(X, f(X), f^{(1)}(X), \dotsc, f^{(m)}(X)) = \tau Q (X, f(X), f^{(1)}(X), \dotsc, f^{(m+1)})(X). \]

We will extensively use the following simple property of the $\tau$ operator.

\begin{proposition}\label{prop: tau product rule}
  Let $\alpha$ be a field element and $\vbeta = (\beta_0,\dots, \beta_i,\dots, \beta_{s-1}) \in \F^s$. Then, for any integer $j$ with $1 \leq j \leq s-m$, \[ \tau^{(j)}[(X-\alpha)Q](\alpha, \vbeta) = j \tau^{(j-1)}(Q)(\alpha, \vbeta). \]
\end{proposition}

\begin{proof}
We prove this by induction on $j$. When $j=1$, the left-hand side is $\tau[(X-\alpha)Q](\alpha, \vbeta)$. From the definition, this equals $[(X-\alpha)\tau Q + Q](\alpha, \vbeta)$. (This can be checked term by term.) The first term $(X-\alpha)\tau Q$ is zero when evaluated at $(\alpha, \vbeta)$ leaving $Q(\alpha, \vbeta)$ as required.

For the induction step, observe similarly that 
\[  
\begin{split}
  \tau^{(j)}[(X-\alpha)Q] & = \tau^{(j-1)} \tau [(X-\alpha)Q] \\
  & = \tau^{(j-1)}[(X-\alpha) \tau Q + Q] \\
  & = \tau^{(j-1)}[(X-\alpha) \tau Q] + \tau^{(j-1)} Q
\end{split}
\]
Evaluating at $(\alpha, \vbeta)$, by the induction hypothesis,
\[ 
\begin{split}
\tau^{(j-1)}[(X-\alpha) \tau Q](\alpha, \vbeta) + \tau^{(j-1)} Q (\alpha, \vbeta) & = (j-1)\tau^{{j-2}}[\tau Q](\alpha, \vbeta) + \tau^{(j-1)} Q (\alpha, \vbeta) \\
& = (j-1)  \tau^{(j-1)} Q (\alpha, \vbeta) +  \tau^{(j-1)} Q (\alpha, \vbeta) \\
& = j  \tau^{(j-1)} Q (\alpha, \vbeta).
\end{split}
\]
\end{proof}

We now define lattices and state the version of Minkowski's theorem for polynomial lattices.

\begin{definition}[polynomial lattice]\label{defn:lattices}
  Let $\calB = \{ \vv_1, \vv_2, \dotsc, \vv_N \}$ be a set of $m$-dimensional vectors of polynomials, that is, $\vv_i \in \F[X]^m$ for all $i$. The \emph{lattice} $\mathcal{L}_\calB$ generated by this set over the ring $\F[X]$ is the set of all linear combinations 
  \[ \mathcal{L}_\calB := \left\{ f_i \vv_1 + f_2 \vv_2 + \dotsb + f_N \vv_N \colon f_i \in \F[X]\right\} .\]
  
  A \emph{basis} of a lattice is a set $\calB$ of vectors of polynomials such that $\calB$ generates the lattice in the sense of the above definition and the vectors in $\calB$ are linearly independent over the field $\F(X)$ of rational functions over $\F$. 
\end{definition}

\begin{theorem}[polynomial version of Minkowski's theorem]\label{thm:minkowski}
Let $\mathcal{L}$ be an $m$-dimensional polynomial lattice. Then, there is a nonzero vector $\vv$ satisfying
\[ \deg \vv \leq \frac{1}{m} \deg \det \mathcal{L} \] where $\deg \vv$ is the max-degree norm and $\det \mathcal{L}$ is the determinant of the lattice basis.
\end{theorem}

A proof can be found in \cite{GoyalHKS2024}.

\begin{theorem}[{\cite[Theorem 2.1]{Alekhnovich2005},\cite[Theorem 9]{GuptaSSV2012}}]\label{thm: alekhnovich shortest vector}
  Let $B$ be a set of $N$ vectors of dimension $m$. Let $d$ be the maximal degree of the polynomials which are entries of the vectors in $B$. Then, there is an algorithm \ShortVector that finds the shortest vector in $\mathcal{L}_B$, the lattice generated by $B$, in time $\tilde{O}(d(N + m)^\omega)$ where $\omega$ is the exponent in the complexity of matrix multiplication.
\end{theorem}

Here ``shortest vector'' is in the max-degree norm. See \cite{GoyalHKS2024} for details.

We will also use the following subroutine off the shelf, for fast Hermite interpolation and for polynomial Chinese remainders.

\begin{theorem}[{\cite[Algorithm 10.22]{GathenG-MCA}}]\label{thm: fast chinese remainder}
Let $m_1(X), m_2(X), \dotsc, m_r(X)$ be univariate polynomials in $\F[x]$ with pairwise GCD equal to 1. Let $v_1(X), v_2(X), \dotsc, v_r(X) \in \F[X]$ such that $\deg v_i < \deg m_i$ for all $i \in [r]$. Then, we can find the unique polynomial $f(X) \in \F[X]$ of degree less than $d = \sum_i \deg m_i$ such that $f \equiv v_i \mod m_i$ for all $i \in [r]$, in time $O(d \poly \log(d))$.
\end{theorem}

\section{Fast construction of differential equation}\label{sec:technical}

As in Guruswami-Wang, our first step is to construct an ``explanation'' polynomial for the received word. In this section, we show that can be done in near-linear time. Below is the main theorem for this section.

\begin{theorem}\label{thm: Q exists}
Let $R = \left(\alpha, \beta_{j}^{(0)},\beta_{j}^{(1)}, \dotsc, \beta_{j}^{(s-1)}\right)_{\alpha \in S, j \in [\ell]}$ be a received word and let $m\leq s-1$ be a parameter. 

Then, for $D \leq n\ell(s-m)/m + n\ell$, there exists a non-zero polynomial $Q(X, Y_0, Y_1, \dotsc, Y_{m})$ of the form $Q = \tilde{Q}(X) + \sum_i Q_i(X)\cdot Y_i$ of $X$-degree at most $D$ with the following property: if $f(X) \in \F_{<k}[X]$ is such that for at least $(D+k)/(s-m)$ values of $\alpha \in S$, we have 
\[ \left(f(\alpha), f^{(1)}(\alpha), \dotsc, f^{(s-1)}(\alpha)\right) = \left(\beta_j^{(0)}, \beta_j^{(1)}, \dotsc, \beta_j^{(s-1)}\right) \] 
for some $j \in [\ell]$, then, 
\begin{equation}\label{eq:fclose}
  Q\left(X, f(X), f^{(1)}(X), \dotsc,  f^{(m)}(X)\right) \equiv 0 . 
\end{equation}
Furthermore, there is a deterministic algorithm that takes as input $R$ and returns $Q$ as a list of coefficients in time $\tilde{O}(n \poly(s+m+\ell))$ --- specifically, $\tilde{O}(n [\ell(s-m)^2 + m^2(s-m) +(s-m)m^\omega ])$.
\end{theorem}

A natural way to ensure that the polynomial $Q$ satisfies \eqref{eq:fclose}, is to ensure that for each $\alpha \in S$, we have the following conditions
\begin{equation}\label{eq:tauconds}
 \tau^{(t)}(Q)\left(\alpha,\beta_j^0,\beta_j^1,\dotsc,\beta_j^{m+t}\right) =0, \qquad \forall t \in [s-m], \ j \in [\ell].
\end{equation}
This will ensure that every close enough codeword satisfies \eqref{eq:fclose} (see \cref{lem: close enough codewords satisfy eqn} for exact details). We will refer to these conditions as the "$\tau$-conditions". The set of polynomials $Q$ satisfying \eqref{eq:tauconds} are closed under addition and multiplication by a polynomial in $\F[X]$. In other words, they form a $\F[X]$-module (or equivalently a lattice with entries from the univariate ring $\F[X]$). Here, it will be convenient to view the polynomial $Q = \tilde{Q}(X) + \sum_i Q_i(X)\cdot Y_i$ as an $(m+2)$-dimensional vector of the form $(\tilde{Q}, Q_0,Q_1,\dotsc,Q_{m}) \in \F[X]^{m+2}$ where the first component $\tilde{Q}$ is the $Y$-free part of $Q$ and the subsequent components are the coefficients of $Y_0, Y_1, \dotsc, Y_{m}$ respectively. To prove \cref{thm: Q exists}, we will first construct a nice-basis for this lattice and then use Minkowski's theorem to show that there exists a short vector in this lattice. The following proposition gives this nice-basis for a sub-lattice of the lattice of polynomials satisfying the "$\tau$-conditions" and use Minkowski's theorem on this sub-lattice to show a low degree polynomial within this sub-lattice itself. 

\begin{proposition}\label{prop:nice basis}
Let $\ell$ be the list-size and $R = \left(\alpha, \beta_{j}^{(0)},\beta_{j}^{(1)}, \dotsc, \beta_{j}^{(s-1)}\right)_{\alpha \in S, j \in [\ell]}$ be the received word and let $m\leq s-1$ be a parameter. 

Then, there exist $m+2$ polynomials in $\F[X,Y_0,\dots,Y_{m}]$ satisfying the $\tau$-conditions \eqref{eq:tauconds} for all $\alpha \in S$ of the following type
\begin{align*}
\tilde{\vB} &= \prod_{\alpha \in S}(X-\alpha)^{s-m}, \\
\vB_i &= Y_i\cdot\prod_{\alpha \in S}(X-\alpha)^{s-m}, &&\forall i, \text{ such that }0 \leq i \leq \ell -2\\
\vB_{i} &= \tilde{C_i}(X) + \sum_{r=0}^{i-1}Y_r\cdot C_{i,r}(X) + Y_i\cdot \prod_{\alpha \in S}(X-\alpha)^{\ell-1}, &&\forall i, \text{ such that }\ell -1 \leq i \leq m.
\end{align*}
for some polynomials $\tilde{C_i}, C_{i,r} \in \F[X]$ for $\ell -1 \leq i \leq m$ and $0 \leq r \leq i-1$.
Furthermore, there is a deterministic algorithm that takes as input $R$ and returns these $(m+2)$-polynomials in time $\tilde{O}(n [\ell(s-m)^2 + m^2(s-m)])$.
\end{proposition}

The equivalent proposition of \cref{prop:nice basis} in the list-decoding case was easier to obtain since the conditions for all the evaluation points (i.e., $\alpha \in S$) are identical. However, this is not the case when the list-size $\ell$ is more than 1. In order to address this, we first show that a version of \cref{prop:nice basis} can be proved for each $\alpha \in S$ and then combine these different bases for the different $\alpha$ into a single basis that works for all the $\alpha$. The following proposition is the version of \cref{prop:nice basis} for a single $\alpha \in S$ (which is proved in \cref{sec:single location}).
\begin{proposition}\label{prop:single location}
Let $\alpha \in S$ and $\ell$ be the list-size and let $m\le s-1$ be a parameter.

Given $\ell$ possible evaluations (i.e, $\left(\beta_{j}^{(0)},\beta_{j}^{(1)}, \dotsc, \beta_{j}^{(s-1)}\right)$ for each $j \in [\ell]$) at the point $\alpha$, 
There, exist $m+2$ polynomials in $\F[X,Y_0,\dots,Y_{m}]$ satisfying the $\tau$-conditions \eqref{eq:tauconds} for this specific $\alpha \in S$ of the following type
\begin{align*}
\tilde{\vB}_\alpha &= (X-\alpha)^{s-m}, \\
\vB_{\alpha,i} &= Y_i\cdot(X-\alpha)^{s-m}, &&\forall i, \text{ such that }0 \leq i \leq \ell -2\\
\vB_{\alpha,i} &= \tilde{C_i}(X) + \sum_{r=0}^{i-1}Y_r\cdot C_{i,r}(X) + Y_i\cdot (X-\alpha)^{\ell-1}, &&\forall i, \text{ such that }\ell -1 \leq i \leq m.
\end{align*}
for some polynomials $\tilde{C_i}, C_{i,r} \in \F[X]$ for $\ell -1 \leq i \leq m$ and $0 \leq r \leq i-1$. Furthermore, there is a deterministic algorithm that takes as input $\left(\beta_{j}^{(0)},\beta_{j}^{(1)}, \dotsc, \beta_{j}^{(s-1)}\right)_{j \in [\ell]}$  and returns these $(m+2)$ polynomials in time $\tilde{O}(\ell(s-m)^2)$.
\end{proposition}

We then show in \cref{sec:combining bases} that the bases for two disjoint sets $U,V$ of evaluation points (i.e., $U,V \subseteq S$) can be combined into a basis for the set $W = U \cup V$ in time approximately $\tilde{O}(\max(|U|,|V|))$. We use this procedure recursively over the bases constructed via \cref{prop:single location} for each $\alpha \in S$ to obtain a basis that works simultaneously for all $\alpha \in S$ to yield \cref{prop:nice basis}.

\subsection{Construction of basis for a single evaluation point}\label{sec:single location}

In this section, we will fix an evaluation point $\alpha \in S$ and construct a set of $m+2$ linearly independent polynomials (over the ring $\F[X]$) as specified in \cref{prop:single location}. More precisely, we want to show that there exist $m+2$ polynomials of the following type that satisfy the $\tau$-conditions for $\alpha$. Since $\alpha$ will be fixed throughout this section, we will drop $\alpha$ from the subscript to declutter the notation.
\begin{align*}
\tilde{\vB}&= (X-\alpha)^{s-m}, \\
\vB_{i} &= Y_i\cdot(X-\alpha)^{s-m}, &&\forall i, \text{ such that }0 \leq i \leq \ell -2\\
\vB_{i} &= \tilde{C_i}(X) + \sum_{r=0}^{i-1}Y_r\cdot C_{i,r}(X) + Y_i\cdot (X-\alpha)^{\ell-1}, &&\forall i, \text{ such that }\ell -1 \leq i \leq m.
\end{align*}
for some polynomials $\tilde{C_i}, C_{i,r} \in \F[X]$ for $\ell -1 \leq i \leq m$ and $0 \leq r \leq i-1$.

To this end, given a polynomial $\vB$ of the above form and any $\vbeta=(\beta^{(0)},\beta^{(1)},\dotsc,\beta^{(s-1)})$ we will find it convenient to use the following notation for the vector in $\F^{s-m}$ composed of evaluations of iterated $\tau$-operator of the polynomial $\vB$ at the point $(\alpha,\vbeta)$:
\[ \vtau(\vB,\vbeta) := \tau^{(<s-m)}(\vB)(\alpha,\vbeta) =
\begin{bmatrix}
  \vB(\alpha, \vbeta) \\
  \tau(\vB)(\alpha, \vbeta) \\
  \tau^{(2)}(\vB)(\alpha, \vbeta) \\
  \vdots \\
  \tau^{(s-m-1)}(\vB)(\alpha, \vbeta) \\
\end{bmatrix} .
\]
The linearity of the $\tau$ operator shows that the function $\vtau(\cdot,\cdot)$ is linear it its first argument.
\begin{proof}[Proof of \cref{prop:single location}]
In the notation defined above, we want to show that $\vtau(\vB,\vbeta)=\vzero_{s-m}$ for all $\vbeta \in \{ \vbeta_j \colon j \in [\ell]\}$ and each $\vB \in \{ \tilde{\vB}, \vB_0,\vB_1,\dotsc,\vB_{m}\}$. Clearly, this is true when $\vB \in \{\tilde{\vB}, \vB_0,\vB_1,\dotsc,\vB_{\ell-2}\}$. In the rest of the proof, we will construct polynomials $\tilde{C_i}, C_{i,r} \in \F[X]$, such that this is also true for the polynomials $\vB = \{\vB_{\ell-1},\vB_{\ell},\dotsc,\vB_{m}\}$. We will prove this by induction on $\ell$, the list size. 
\begin{description}
\item[Base case $\ell=1$:] This is the list decoding setting, where we have only $\beta$ in the list corresponding to $\alpha$, namely $\vbeta_0 = (\beta^{(0)}_0,\beta^{(1)}_0,\dotsc,\beta^{(s-1)}_0)$. The proof in this case is similar to the corresponding step in \cite{GoyalHKS2024}, which is obtained by Hermite interpolation. For each $0\le i \le m$, we define the polynomial $\tilde{C_i}$ to be the unique $(s-m-1)$-degree polynomial in $\F[X]$ such that for all $j \in [s-m]$, we have 
\[{\tilde{C_i}}^{(j)}(\alpha)=-\beta_{0}^{(i+j)}.\] 

The remaining $C_{i,r}(X)$ are set to zero for all $0\le i \le m$ and $r \in [i]$. In other words, the $i^{th}$ polynomial $\vB_i :=Y_i + \tilde{C}_i(X)$ for $\ell-1=0\leq i \leq m$. $\vB_i$ clearly satisfies the $\tau$-conditions \eqref{eq:tauconds} for $\alpha$.

\item[Induction Step $\ell \geq 2$:] By induction, let us assume that for the first $(\ell-1)$ list elements $\vbeta_{j}, j \in [\ell-1]$, we have polynomials $\vB'_{i}$ satisfying the $\tau$-conditions for $\alpha$, namely $\vtau(\vB'_i,\vbeta_j) = \vzero$ for all $i$ such that $\ell -2 \leq i \leq m$ and $j \in [\ell-1]$.  We now need to construct polynomials $\vB_i$ for $\ell -1 \leq i \leq m$ such that $\vtau(\vB_i,\vbeta_j)=\vzero$ for all $i$ such that $\ell-1 \leq i \leq m$ and $j \in [\ell]$. Note that the number of polynomials to be constructed in the $\ell$-case is one less than the $(\ell-1)$-case while the number of $\vbeta$'s for which we need $\vtau(\vB_i,\vbeta_j)=\vzero$ is one more.

Fix an $i$ such that $\ell-1 \leq i \leq m$. We will construct the polynomial $\vB_i$ from the polynomial $\vB'_{i-1}$. Consider the polynomial $\vB''_i := \tau\left((X-\alpha)\cdot \vB_{i-1}\right)$. Clearly, this polynomial is of the required form $Y_i \cdot (X-\alpha)^{\ell-1} + \sum_{r \in [i]}Y_r \cdot C_{i,r}(X) + \tilde{C}_i(X)$. By \cref{prop: tau product rule}, we have that $\vtau(\vB''_i,\vbeta_j)=\vzero$ for all $j\in [\ell-1]$. Now, if $\vtau(\vB''_i,\vbeta_{\ell-1})$ is also $\vzero$, we can set $\vB_i : = \vB''_i$ and be done. 

Suppose $\vtau(\vB''_i,\vbeta_{\ell-1}) \neq \vzero$. Then, let $r_* \in [s-m]$ be the least $r$ such that $\tau^{(<r)}(\vB''_i)(\alpha,\vbeta_{\ell-1}) = \vzero_r$ but $\tau^{(r)}(\vB''_i)(\alpha,\vbeta_{\ell-1}) \neq 0$. By \cref{prop: tau product rule}, definition of $\vB''_i$ and the fact that $s-m$ is less than the characteristic of $\F$, we also have $\tau^{(<r)}(\vB'_{i-1})(\alpha,\vbeta_{\ell-1}) = \vzero_r$ but $\tau^{(r)}(\vB'_{i-1})(\alpha,\vbeta_{\ell-1})\neq 0$. Let us consider the $(s-m-r_*)$ polynomials $\{(X-\alpha)^{t} \cdot \vB'_{i-1} \colon t \in [s-m-r_*]\}$. The $t$-th polynomial in this set, namely $(X-\alpha)^{t} \cdot \vB'_{i-1}$ satisfies $\tau^{(<r+t)}((X-\alpha)^{t}\cdot \vB'_{i-1})(\alpha,\vbeta_{\ell-1})=\vzero_{r+t}$ but $\tau^{(r+t)}((X-\alpha)^t\cdot \vB_{i-1})(\alpha,\vbeta_{\ell-1}) \neq 0$. Hence, the $(s-m-r_*)$ vectors $\{\vtau\left((X-\alpha)^t\cdot \vB'_{i-1},\vbeta_{\ell-1}\right) \colon t \in [s-m-r_*]\}$ form a basis for the space of all vectors in $\F^{s-m}$ which are zero in the first $r_*$ coordinates. This space contains the vector $\vtau(\vB''_i,\vbeta_{\ell-1})$ and hence there exist scalars $c_t \in \F$ for $t \in [s-m-r_*]$ such that 
\[\vtau(\vB''_i,\vbeta_{\ell-1}) = \sum_{t \in [s-m-r_*]} c_t \cdot \vtau((X-\alpha)^t \cdot \vB'_{i-1},\vbeta_{\ell-1}).\] We can now set 
\[\vB_i := \vB''_i - \sum_{t \in [s-m-r_*]} c_t \cdot (X-\alpha)^t\cdot \vB'_{i-1}.\] By choice of the scalars $c_t$, we have that $\vtau(\vB_i,\vbeta_{\ell-1}) = \vzero$. Since $\vtau(\vB''_i,\vbeta_j) =\vzero$ and $\vtau(\vB'_{i-1},\vbeta_j)=\vzero$ for all $j \in [\ell-1]$, we also have that $\vtau(\vB_i,\vbeta_{j}) = \vzero$ for all $j \in [\ell-1]$. Combining, we have that $\vtau(\vB_i,\vbeta_{j}) = \vzero$ for all $j \in [\ell]$.

\end{description}
It can be seen that $\vB_i$ is of the required form $Y_i \cdot (X-\alpha)^{\ell-1} + \sum_{r \in [i]}Y_r \cdot C_{i,r}(X) + \tilde{C}_i(X)$ since $\vB''_i$ is of this form and each of the polynomials $(X-\alpha)^t \cdot \vB'_{i-1}$ are of the form $\sum_{r \in [i]} Y_r \cdot C_{i,r}(X) + \tilde{C}_i(X)$.

We turn to the time complexity. For each list element, we may have to solve a system of equations to find the scalars $\{c_t\}$. Observe that this is a triangular system in at most $s-m$ variables, and hence can be solved in $O(s-m)^2$ time; giving an overall total of $\tilde{O}(\ell(s-m)^2)$ time. This concludes the proof of the proposition.
\end{proof}

\subsection{Combining the bases}\label{sec:combining bases}

In the previous section, we constructed a nice basis for each $\alpha \in S$. In this section, we will show how to combine these bases recursively to obtain a single basis that works for all $\alpha \in S$.  To this end, we begin by showing a general procedure to combine the bases of two $\F[X]$-modules to generate the basis for a related $\F[X]$-module. 

Let $S$ be the set of evaluation points. For each $\alpha \in S$, let $p_\alpha(X)$ be an associated monic polynomial of degree at most $(s-m)$ such that for any two distinct $\alpha, \beta \in S$, we have $\gcd(p_\alpha,p_\beta)=1$. We will be setting $p_\alpha(X)= (X-\alpha)^{s-m}$. 

Let $\calU,\calV$ be two disjoint non-empty subsets of $S$. Define the polynomials 
\begin{align*}
P_\calU(X) : = \prod_{\alpha \in \calU} p_\alpha(X), && P_\calV(X):= \prod_{\alpha \in \calV} p_\alpha(X).
\end{align*} Observe that since $\calU \cap \calV = \emptyset$, we have $\gcd(P_\calU,P_\calV)=1$. 

Let $\calM_\calU, \calM_\calV$ be two $(m+2)$-dimensional $\F[X]$-modules\footnote{Here, we identify the polynomial $\tilde{Q} + \sum_{i \in[m]} Y_i \cdot Q_i(X)$ with the $(m+2)$-dimensional vector $(\tilde{Q},Q_0,Q_1,\dotsc,Q_{m})$} with lower-triangular bases $\calB_\calU = \{\tilde{\vU},\vU_0,\vU_1,\dotsc,\vU_{m}\}$ and $\calB_\calV = \{\tilde{\vV}, \vV_0,\vV_1,\dotsc,\vV_{m}\}$ respectively of the following type:
\begin{align*}
\tilde{\vU} &= \tilde{U}(X), & \tilde{\vV} &= \tilde{V}(X)\\
\vU_i & = \tilde{U_i}(X) + \sum_{r =0}^i Y_r\cdot U_{i,r}(X), & \vV_i & = \tilde{V_i}(X) + \sum_{r =0}^i Y_r\cdot V_{i,r}(X), & \forall 0\le i \le m
\end{align*}
Furthermore, suppose these bases $\calB_\calU, \calB_\calV$ behave "nicely" with their respective polynomials $P_\calU, P_\calV$ in the following sense: the diagonal elements $\tilde{U}, U_{i,i}, i \in [m]$ are non-zero factors of the polynomial $P_\calU$ and similarly $\tilde{V}, V_{i,i}, i \in [m]$ are non-zero factors of the polynomial $P_\calV$. 

Let $\calW = \calU \cup \calV$ and $P_\calW = P_\calU \cdot P_\calV$ be the associated polynomial with the set $\calW$. We can construct an lower-triangular basis $\calB_\calW=\{\tilde{\vW},\vW_0,\vW_1,\dotsc,\vW_{m}\}$ which behaves "nicely" with the polynomial $P_\calW$ as follows:
\begin{align*}
\tilde{\vW} &= \tilde{W}(X)\\
\vW_i & = \tilde{W_i}(X) + \sum_{r =0}^i Y_r\cdot W_{i,r}(X), && \forall 0\le i\le m
\end{align*}
where the diagonal elements are given by $\tilde{W} = \tilde{U}\cdot \tilde{V}$ and $W_{i,i} = U_{i,i}\cdot V_{i,i}, 0\le i \le m$ and the non-diagonal elements $W_{i,r}(X), r \in [i]$ are the unique polynomials (by the Chinese remainder theorem) such that 
\[
\begin{split}
W_{i,r} & \equiv U_{i,r} \cdot V_{i,i} \mod P_\calU,\\
W_{i,r} & \equiv V_{i,r} \cdot U_{i,i} \mod P_\calV.
\end{split} \]
This basis satisfies that $\vW_i \equiv V_{i,i}(X)\cdot \vU_i \bmod P_\calU$ and symmetrically, $\vW_i \equiv U_{i,i}(X) \cdot \vV_i \bmod P_\calV$. 

Let $\calM_\calW$ be the $\F[X]$-module generated by this basis $\calB_\calW$. This basis satisfies the following nice property.
\begin{claim}\label{clm: combining bases}
  If $\vR = \tilde{R}(X) + \sum_{i \in [m]} Y_i \cdot R_i(X)$ or equivalently $(\tilde{R},R_0,R_1,\dotsc,R_{m})$ satisfies $\vR \bmod P_\calU \in \calM_\calU$ and $\vR \bmod P_\calV \in \calM_\calV$, then $\vR \bmod P_\calW \in \calM_\calW$.
\end{claim}
\begin{proof}
Since $\vR \bmod P_\calU \in \calM_\calU$ and $\vR \bmod P_\calV \in \calM_\calV$, we have polynomials $\tilde{A}, \tilde{B}, A_i, B_i, 0\le i\le m$ such that
\begin{align*}
\vR \bmod P_\calU & = \tilde{A}(X)\cdot \tilde{\vU} +\sum_{i=0}^m A_i(X) \cdot \vU_i,\\
\vR \bmod P_\calV & = \tilde{B}(X) \cdot \tilde{\vV}+\sum_{i=0}^m B_i(X)\cdot \vV_i. 
\end{align*}
Define polynomials 
\begin{align*}
\tilde{C}(X) &:= \tilde{U}^{-1}\cdot P_\calU^{-1} \cdot \tilde{B} \bmod P_\calV,& \tilde{D}(X) &:= \tilde{V}^{-1}\cdot P_\calV^{-1} \cdot \tilde{A} \bmod P_\calU\\
C_i(X) &:= U_{i,i}(X)^{-1}\cdot P_\calU^{-1}\cdot B_i \bmod P_\calV, & D_i(X) &:= V_{i,i}^{-1}\cdot P_\calV^{-1}\cdot A_i \bmod P_\calU, & i \in [m]
\end{align*}
We claim that \[ \vR = [\tilde{C}\cdot P_\calU+ \tilde{D}\cdot P_\calV]\tilde{\vW}+\sum_{i=0}^m [C_i\cdot P_\calU + D_i\cdot P_V] \vW_i \bmod P_U P_V. \]
Indeed, observe the right-hand side mod $P_\calU$. All the terms involving $P_\calU$ vanish, leaving $\tilde{D}\cdot P_\calV \cdot \tilde{\vW}+\sum_{i=0}^m D_i\cdot P_\calV\cdot \vW_i$. Simplifying using the definition of $\vW$, we have $\tilde{A}\cdot \tilde{\vU} + \sum_{i \in [m]} A_i \cdot \vU_i$, which is $\vR\bmod P_\calU$. The case mod $P_V$ is symmetric and the equivalence follows from the Chinese remainder theorem.
\end{proof}

For our specific setting, the polynomial $p_\alpha(X)$ will be $(X-\alpha)^{s-m}$. For any subset $\calU \subseteq S$ of the evaluation points, we will say that the basis $\calB_\calU = \{\tilde{\vU},\vU_0,\dots,\vU_{m}\}$ is $\calU$-good, if the following two properties are satisfied.
\begin{enumerate}
\item \label{item:diagonal} The diagonal elements of the basis satisfy
\begin{align*}
\tilde{U}(X) &:= \prod_{\alpha \in \calU} (X-\alpha)^{s-m},\\
U_{i,i}(X) &:= \prod_{\alpha \in \calU} (X-\alpha)^{s-m}, \qquad 0 \leq i \leq \ell -2\\
U_{i,i}(X) &:= \prod_{\alpha \in \calU} (X-\alpha)^{\ell-1}, \qquad \ell-1 \leq i \leq m
\end{align*}
where $\ell$ is the list size.
\item \label{item:taucond} Every basis element $\vU \in \calB_\calU$ satisfies the $\tau$-conditions \eqref{eq:tauconds} for every $\alpha \in \calU$.
\end{enumerate}

Clearly, the basis generated by \cref{prop:single location} when $\calU = \{\alpha\}$ is a singleton set is $\{\alpha\}$-good. The following claim shows that if $\calB_\calU, \calB_\calV$ are both good, so is $\calB_\calW$.
\begin{claim}\label{clm:good}
  Let $\calU,\calV \subseteq S$ be two disjoint non-empty subsets of the set of evaluation points $S$. If $\calB_\calU$ is $\calU$-good and $\calB_\calV$ is $\calV$-good, then $\calB_\calW$ is $\calW$-good.

  Furthermore, there is a deterministic algorithm that when given the bases $\calB_\calU$ and $\calB_\calV$ outputs the basis $\calB_\calW$ in time $\tilde{O}(\max\{|\calU|,|\calV|\}m^2(s-m))$.
\end{claim}
\begin{proof}
\cref{item:diagonal} follows from how the diagonal elements of the basis $\calB_\calW$ are defined. The runtime bound follows from the time required for Chinese Remainder theorem (\cref{thm: fast chinese remainder}): for the $O(m^2)$ off-diagonal elements $W_{i, r}$, we carry out a CRT of polynomials of degree at most $\max\{|\calU|,|\calV|\}(s-m)$.

We now prove \cref{item:taucond}. Let $\vW \in \calB_\calW$. If $\vW = \tilde{\vW}$, then $\vW = \tilde{\vV} \cdot \tilde{\vU} = \prod_{\alpha \in \calW}(X-\alpha)^{s-m}$ and hence satisfies the $\tau$-conditions for all $\alpha \in \calW$. Suppose $\vW = \vW_i$ for some $0\le i \le m$. Let $\alpha \in \calW$. Suppose $\alpha \in \calU$. By definition of $\vW_i$, we have $\vW_i \equiv V_{i,i} \cdot \vU_i \bmod P_\calU$ and is hence of the form $V_{i,i}\cdot \vU_i + \prod_{\alpha \in \calU} (X-\alpha)^{s-m}\cdot \vR$ for some polynomial $\vR = \tilde{R}+\sum_{i=0}^mY_i \cdot R_i(X)$. Since $\calB_\calU$ is $\calU$-good, $\vU_i$ satisfies the $\tau$-conditions with respect to $\alpha$. Hence, so does $V_{i,i}\cdot \vU_i$ and $\vW_i$. The case when $\alpha \in \calV$ is similar.
\end{proof}

We are now ready to prove \cref{prop:nice basis}.
\begin{proof}[Proof of \cref{prop:nice basis}]
Let $\calB_S:=\{\tilde{\vB}, \vB_0, \vB_1, \dotsc, \vB_{m}\}$ be obtained by the following process: Divide $S$ into two halves $\calU$ and $\calV$, and recursively compute bases $\calB_\calU$ and $\calB_\calV$ for them, and then combine them as indicated earlier in this section. For a singleton set, we use the basis given by \cref{prop:single location}.

It follows from \autoref{clm:good}, that since the bases for the singleton sets are good, so are all the bases that are recursively constructed. Thus, the basis $\calB_S$ has all the properties mentioned in \cref{prop:nice basis}. The only thing left to be verified is the running time. 

As for the running time, let $T(n)$ be the running time where $n$ is the size of the set $S$. We split it into two instances of size $n/2$, each of which can be solved in time $T(n/2)$. \autoref{clm:good} states that the bases for the two instances can be combined in time $\tilde{O}(n m^2 (s-m))$. Finally, \cref{prop:single location} states that the base case when the set is a single point requires time $ \tilde{O}(\ell(s-m)^2)$. Putting all this together, we have  $T(n) = 2T(n/2) + \tilde{O}(n m^2 (s-m))$ and $T(1) = \tilde{O}(\ell(s-m)^2)$. Solving, we get $T(n) = \tilde{O}(n [\ell(s-m)^2 + m^2(s-m)])$. 
\end{proof}

\subsection{Proof of \cref{thm: Q exists}}

Before we prove the main theorem of the section, we have to connect the $\tau$-conditions to the desired property, of close enough codewords satisfying the differential equation. 

\begin{lemma}\label{lem: close enough codewords satisfy eqn}
  Let $Q(X, Y_0, Y_1, \dotsc, Y_{m}) \in \calM_S$ have $X$-degree at most $D \leq n\ell(s-m)/m + n\ell$ and $Y_i$-degree at most 1 for every $Y_i$. Let $f(X) \in \F_{<k}[X]$ be such that for $(D+k)/(s-m)$ values of $\alpha \in S$, $(f(\alpha), f^{(1)}(\alpha), \dotsc, f^{(s-1)}(\alpha)) = (\beta_j^{(0)}, \beta_j^{(1)}, \dotsc, \beta_j^{(s-1)})$ for some $j \in [\ell]$. Then, $Q(X, f(X), f^{(1)}(X), \dotsc,  f^{(s-1)}(X)) \equiv 0 $.
\end{lemma}

\begin{proof}
Let $P(X) = Q(X, f(X), f^{(1)}(X), \dotsc,  f^{(s-1)}(X))$. Since $f$ has degree at most $k$ and $Q$ is linear in the $Y$-variables, $P(X)$ has degree at most $D+k$. The definition of the  $\tau$ operator is set up such that $\tau^{(i)}(Q)(X, f(X), f^{(1)}(X), \dotsc,  f^{(s-1)}(X)) = \frac{d^iP}{dX^i}$. This combined with \cref{prop:nice basis} implies that for an agreement point $\alpha$,
\[ \frac{d^iP}{dX^i}(\alpha) = [\tau^{(i)}(Q)](\alpha, f(\alpha), f^{(1)}(\alpha), \dotsc,  f^{(s-1)}(\alpha)) = \tau^{(i)}(Q)(\alpha,\beta_j^{(0)}, \beta_j^{(1)}, \dotsc, \beta_j^{(s-1)}) = 0\] 
for any $i \in [s-m]$. That is, at these points, $P$ vanishes with multiplicity at least $(s-m)$. Therefore, if the number of such agreements is more than $(D+k)/(s-m)$, $P$ must be identically zero.
\end{proof}

We are now ready to prove the main theorem of the section.

\begin{proof}[Proof of \cref{thm: Q exists}]

  We will take $Q$ to be a polynomial in the $\F[X]$-span of the polynomials $\tilde{\vB}, \vB_0, \vB_1, \dotsc, \vB_{m}$ given by \cref{prop:nice basis}. Since each of them satisfies the $\tau$ conditions, as given by the proposition, so will $Q$. In addition, $Q$ will have the form $Q = \tilde{Q}(X) + \sum_i Q_i(X)\cdot Y_i$, inherited from the basis polynomials.

  We now have to bound the degree of the coefficients. Following our previous notation, let $\calM_S$ be $\spn \{\tilde{\vB}, \vB_0, \vB_1, \dotsc, \vB_{m} \}$. By \cref{thm:minkowski}, $\calM_S$ contains a polynomial of degree at most $(\deg \det \calM_S)/(m+2)$. Because of the lower triangular structure, to find $\deg \det \calM_S$, we need only multiply the degrees of the terms on the diagonal.

  The first $\ell$ terms on the diagonal are $\prod_{\alpha \in S}(X-\alpha)^{s-m}$, and the remaining $m+2-\ell$ are $\prod_{\alpha \in S}(X-\alpha)^{\ell-1}$. This gives $\deg \det \calM_S = n\ell(s-m) + (m+2-\ell)n(\ell-1) \leq n\ell(s-m)+n(m+2)\ell$. Applying the Minkowski statement gives the required bound.

  The close enough codewords statement is given by \cref{lem: close enough codewords satisfy eqn}. This gives all the required properties of $Q$.

  The running time is the time to construct the basis followed by the time to find a short vector in the lattice. From \cref{prop:nice basis} and \cref{thm: alekhnovich shortest vector}, we get a running time bound of $\tilde{O}(n [\ell(s-m)^2 + m^2(s-m) +(s-m)m^\omega ])$.
\end{proof}

\section{List recovery algorithm}\label{sec:proof of main thm}

Having constructed the differential equation, we now have to solve it in near-linear time. We will be using subroutines off the shelf for this. We mention the relevant algorithms before analyzing the overall list recovery algorithm.

\subsection{Fast algorithms for solving differential and functional equations}  

We will invoke the following algorithms for solving differential equations from \cite{GoyalHKS2024}. (Their paper also contains an analogous result for functional equations, for FRS codes.)

\begin{theorem}[{\cite[Theorem 5.1]{GoyalHKS2024}}]\label{thm: fast DE solver}
Let $\F$ be a finite field of characteristic greater than $d$ or zero, and let \[Q(X, Y_0, \ldots, Y_m) = \tilde{Q}(X) + \sum_{i = 0}^m Q_i(x)Y_i\] be a non-zero polynomial with $X$-degree at most $D$. Then, the set of polynomials $f(X) \in \F[X]$ of degree at most $d$ that satisfy 
\[
Q\left(X, f, f^{(1)} \ldots, f^{(m)}\right) \equiv 0
\]
is contained in an affine space dimension at most $m$. 

Furthermore, there is a deterministic algorithm \FastDESolver that when given $Q$ as an input via its coefficient vector, and the parameter $d$, performs at most $\tilde{O}((D+d)m^4)$ field operations and outputs a basis for this affine space.
\end{theorem}

\begin{remark}
  The \cite[Theorem 5.1]{GoyalHKS2024} paper presents a complexity of $\tilde O((D+d)\poly(m))$. On closer examination, their recursion contains a factor of at most $m^4$ as they solve for $m$ different basis elements separately and each is solved using \cite[Algorithm 1]{GoyalHKS2024}. The time complexity for this step is analyzed in \cite[Lemma 5.8]{GoyalHKS2024} which their analysis shows has a dependence of at most $O(m^3)$.
\end{remark}

These algorithms output a subspace containing the codewords. We use the ``Prune'' subroutine introduced by \cite{KoppartyRSW2023}, which was simplified by \cite{Tamo2024}, to obtain the codewords. The list size bounds we present below are due to \cite{Tamo2024}.

\begin{theorem}[{\cite[Lemma 3.1]{KoppartyRSW2023}}, {\cite[Theorem 4.5]{Tamo2024}}]\label{thm:prune}
  For every $\epsilon>0, \ell \in \mathbb N$, all integers $s>\frac{16\ell}{\epsilon^2}$, degree parameter $k$, block length $n$ and field $\FF$ of characteristic zero or greater than $k$, the following is true.

  Define \[ L(\delta, \epsilon, \ell, r) =  \left( \frac{\ell}{\epsilon}\right)^{O(\frac{1+\log (\ell)}{\epsilon})}. \]
Then, for any $\gamma >0$, there is a randomized algorithm \Prune that when given as input sets $S\subseteq \FF$ and $E_\alpha\subseteq \FF^s$, with $|E_\alpha|\le \ell$ for every $\alpha \in S$ and $|S|=n$, and a basis for an affine space $\mathcal A$ of dimension at most $4\ell/\epsilon$, outputs \[\calL = \{f(X)\in \FF_{<k}[X]\cap \mathcal A\mid |\{\alpha \in S\mid (f(\alpha), f^{(1)}(\alpha),\cdots, f^{(s-1)}(\alpha))\in E_\alpha\}|>(R+\epsilon)n\}\] with probability at least $1-\gamma$ in time $\tilde{O}(n) \poly(\log q, s, L(\delta, \epsilon, \ell, r), \log (1/\gamma))$ and it is guaranteed that, 
  \[ \abs{\calL} \leq L(\delta, \epsilon, \ell, r)\ . \]

\end{theorem}

\subsection{Algorithm and proof of main theorem}

The final algorithm proceeds in $3$ steps: finding the differential equation, solving it, and pruning. Before restating and proving the main theorem, we state separately here the combination of the first two steps, as the time complexity in those parts is our main contribution.

\begin{theorem} \label{thm: Alekhnovich+GHKS24}
  For all integers $\ell, m, s\in \mathbb N$ with $m<s$, degree parameter $k$, block length $n$ and field $\FF$ of characteristic zero or greater than $k$, the following is true.
  
  There is a deterministic algorithm, that when given as input sets $S\subseteq \FF$ and $E_\alpha\subseteq \FF^s$ with $|E_\alpha|\le \ell$ for every $\alpha \in S$ and $|S|=n$, runs in time $O(n\log |\FF|(\ell (s^2+sm^3)+ \frac{k}{n}m^4)\polylog(n, s, \log\log |\FF|))$ and outputs the basis of an affine space $\mathcal A\subseteq \FF[X]$ of dimension at most $m$ such that $\mathcal A$ contains all polynomials $f(X)$ of degree less than $k$ such that for at at least $(\frac{n\ell+k}{n(s-m)}+\frac{\ell}{m})$ fraction of $\alpha \in S$, \[(f(\alpha),\cdots, f^{(s-1)}(\alpha))\in E_\alpha \ .\]
\end{theorem}
\begin{proof}
We begin by describing the algorithm.

\textbf{Input: } $R = \{(\alpha_i, \beta_j^{(0)},  \beta_j^{(1)}, \dotsc,  \beta_j^{(s-1)}) \}_{i \in [n], j \in [\ell]}, k \in \Z$

\textbf{Task: } Return the basis of an affine space $\mathcal A$ of dimension at most $m$ containing all $f \in \F_{< k}[X]$ such that for $n\ell/m + (n\ell+k)/(s-m)$ values of $i$, \[(f(\alpha_i), f^{(1)}(\alpha_i), \dotsc, f^{(s-1)}(\alpha_i)) = (\beta_j^{(0)},  \beta_j^{(1)}, \dotsc,  \beta_j^{(s-1)})\] for some $j \in [\ell]$.

\begin{enumerate}
  \item Construct a multivariate polynomial $Q$ as per \cref{thm: Q exists}. 
  \item Solve the resulting differential equation using the algorithm in \cref{thm: fast DE solver}, to obtain an affine space $A$ containing all the codewords.
\end{enumerate}

Observe that by \Cref{thm: Q exists}, we can construct a multivariate polynomial $Q(X, Y_0, \cdots, Y_m)$ of the form $Q=\tilde Q(X)+\sum_i Q_i(X)\cdot Y_i$ of $X$-degree at most $D\le \tfrac{n\ell(s-m)}{m}+n\ell$ with the following property: if $f(X)\in \FF_{<k}[X]$ is such that for at least $\frac{1}{s-m} \left( \frac{n\ell(s-m)}{m}+n\ell+k \right) = \frac{n\ell}{m}+\frac{n\ell+k}{s-m}$ values of $\alpha \in S$, we have \[(f(\alpha),\cdots, f^{(s-1)(\alpha)}\in E_\alpha)\ , \] then, \[Q(X, f(X), f^{(1)}(X),\cdots, f^{(m)}(X))\equiv 0 \]
in time $\tilde O(n\ell ((s-m)^2+(s-m)m^\omega))$ which is within our time complexity.

Now, \Cref{thm: fast DE solver} outputs the basis of the affine space of dimension at most $m$ of the polynomials of degree $<k$ which are solutions of the above differential equation in time $\tilde O((D+k)m^4)$. Substituting $D\le \tfrac{n\ell(s-m)}{m}+n\ell$, the time complexity is within the claimed range.
\end{proof}

We are finally ready to prove the main theorem. We first recall the theorem statement.

\mainthm*
\begin{proof}

We will first set the parameters. Recall that the agreement fraction for which \cref{thm: Q exists} works is \[ \frac{n\ell+k}{n(s-m)} + \frac{\ell}{m}. \]

The rate $R$ is $k/ns$. For the above to approximately equal $R + \epsilon$, we set $s_0 = 16\ell/\epsilon^2+\frac{4\ell}{\epsilon}$ and $m = 4\ell/\epsilon$ with $s>s_0$. Thus, \[\frac{n\ell+k}{n(s-m)} + \frac{\ell}{m}\le \frac{\ell}{(s-m)}+\frac{k}{n(s-m)}+\frac{\epsilon}{2}\le\frac{\epsilon^2}{16}+\frac{\epsilon}{4}+R\cdot\left(\frac{s}{s-m}\right)\le R+\frac{\epsilon\cdot R}{4}+\frac{\epsilon}{4}+\frac{\epsilon^2}{16}\] 
Since $\epsilon\le 1-R$ and thus $\epsilon^2\le \epsilon (1-R)$, we get that \[ \frac{n\ell+k}{n(s-m)} + \frac{\ell}{m}< R + \epsilon/2. \]

The algorithm is given below.
  
\textbf{Input: } $R = \{(\alpha_i, \beta_j^{(0)},  \beta_j^{(1)}, \dotsc,  \beta_j^{(s-1)}) \}_{i \in [n], j \in [\ell]}, k \in \Z$

\textbf{Task: } Return all $f \in \F_{< k}[X]$ such that for $n\ell/m + (n\ell+k)/(s-m)$ values of $i$, \[(f(\alpha_i), f^{(1)}(\alpha_i), \dotsc, f^{(s-1)}(\alpha_i)) = (\beta_j^{(0)},  \beta_j^{(1)}, \dotsc,  \beta_j^{(s-1)})\] for some $j \in [\ell]$.

\begin{enumerate}
  \item Find an affine space of dimension at most $m$ which contains all the required codewords using \Cref{thm: Alekhnovich+GHKS24}.
  \item Using the algorithm in \cref{thm:prune}, obtain a list $\calL$ of codewords and return it.
\end{enumerate}

Let $f$ be a polynomial such that for $n\ell/m + (n\ell+k)/(s-m)$ values of $i$, \[(f(\alpha_i), f^{(1)}(\alpha_i), \dotsc, f^{(s-1)}(\alpha_i)) = (\beta_j^{(0)},  \beta_j^{(1)}, \dotsc,  \beta_j^{(s-1)})\] for some $j \in [\ell]$. By \cref{thm: Alekhnovich+GHKS24}, $f$ should be contained in the output affine space. The algorithm in \cref{thm:prune} finds $f$ with high probability.

We saw in \cref{thm: Alekhnovich+GHKS24} that step $1$ can be done in $\tilde{O}(n \poly(s+m+\ell))$ time. Combining with the runtime of \Cref{thm:prune}, we get a runtime of $\tilde O\left(n \poly(s, \ell, (\ell/\epsilon)^{\frac{1+\log \ell}{\epsilon}}) \right)$.

\end{proof}

\addcontentsline{toc}{section}{References}
{\small 
  \bibliographystyle{prahladhurl}
    \bibliography{fastrecovery-bib}
}

\end{document}